\documentclass[%
 reprint,
superscriptaddress,
nofootinbib,
amsmath,
amssymb,
aps,
]{revtex4-2}
\usepackage{subcaption} 

\usepackage{graphicx}
\usepackage{dcolumn}
\usepackage{bm}
\usepackage[normalem]{ulem}
\usepackage{subcaption}
\usepackage{cancel}
\usepackage{hyperref}
\usepackage{cleveref}
\crefname{figure}{Fig.}{Figs.}
\crefname{equation}{Eq.}{Eqs.}
\crefname{section}{Sec.}{Secs.}
\crefname{appendix}{App.}{Apps.}
\crefname{table}{Tab.}{Tabs.}
\usepackage{amsmath, amsfonts, amsthm, mathtools, amssymb, amsbsy} 
\usepackage{xcolor}

\usepackage{comment}

\usepackage{placeins}

\begin{document}
\preprint{APS/123-QED}

\title{Theory of In-Plane-Magnetic-Field-Dependent Excitonic Spectra in Atomically Thin Semiconductors}

\author{Michiel Snoeken}
\affiliation{Nichtlineare Optik und Quantenelektronik, Institut f\"ur Physik und Astronomie (IFPA), Technische Universit\"at Berlin, 10623 Berlin, Germany}
\author{Paul Steeger}
\author{Robert Schmidt}
\author{Steffen Michaelis de Vasconcellos}
\author{Rudolf Bratschitsch}
\affiliation{Institute of Physics and Center for Nanotechnology, University of Münster, 48149 Münster, Germany}
\author{Andreas Knorr}
\affiliation{Nichtlineare Optik und Quantenelektronik, Institut f\"ur Physik und Astronomie (IFPA), Technische Universit\"at Berlin, 10623 Berlin, Germany}
\author{Henry Mittenzwey}
\affiliation{Nichtlineare Optik und Quantenelektronik, Institut f\"ur Physik und Astronomie (IFPA), Technische Universit\"at Berlin, 10623 Berlin, Germany}
\affiliation{Institut für Theoretische Physik und Zentrum für Materialforschung, Justus-Liebig-Universität Gießen, 35392 Gießen, Germany}

\date{\today}

\begin{abstract}
The linear absorption spectrum of excitons in TMDC monolayers under the influence of an in-plane magnetic field is theoretically studied. We demonstrate that in-plane magnetic fields induce a hybridization between spin-bright and spin-dark exciton transitions, resulting in a brightening of spin-dark excitons. We analytically investigate spectral features including resonance energy shifts, broadening and amplitudes ratios. In particular, for a MoSe$_2$ monolayer with linewidths dominated by reradation, we find a complex interplay of dark-bright splitting and linewidth difference of both involved spin-bright and spin-dark excitons.
\end{abstract}

\maketitle

\section{\label{sec: Introduction} Introduction}

Transition metal dichalcogenides (TMDC) monolayers are atomically thin semiconductors of the $MX_2$ type, where $M=\{\text{Mo},\text{W}\}$ and $X = \{\text{S},\text{Se}\}$. The screening of the Coulomb interaction compared to their bulk forms is drastically reduced, which results in large excitonic binding energies of a few hundreds of meV. As a consequence, excitons dominate the linear optical response of such materials in the visible region \cite{RevModPhys.90.021001,chernikov2014exciton,qiu2013optical}.

\begin{figure}
    \centering
    \includegraphics[width=1\linewidth]{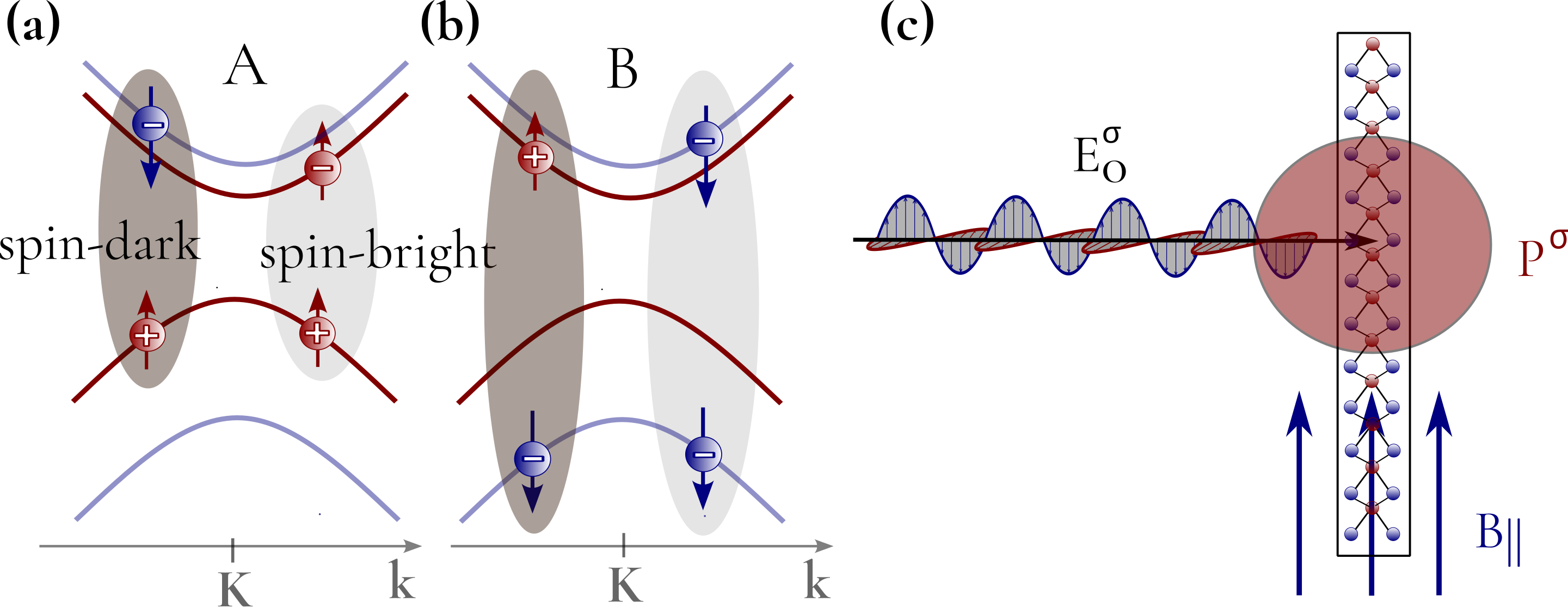}
    \caption{(a) Spin-bright (light-gray ellipse) and spin-dark (dark-gray ellipse) excitonic transitions of the A-exciton considered in our model at the $K$-valleys. (b) Corresponding spin-bright and spin-dark transitions of the B-exciton. (c): An optical field $E_0^{\sigma}$ induces a polarization $P^{\sigma}$ in a TMDC monolayer in the presence of an in-plane magnetic field $B_{\parallel}$.}
    \label{fig:setting}
\end{figure}

Due to strong spin-orbit-interaction-induced band splitting \cite{Kormanyos_2015,PhysRevX.4.011034}, the valley and spin of the electrons in TMDC monolayers provide two degrees of freedom, cf.\ \cref{fig:setting}(a)--(b), which can be optically addressed, giving rise to possible applications in valleytronics \cite{tyulnev2023valleytronics} and spintronics \cite{vzutic2004spintronics}. As depicted in \cref{fig:setting}(a)--(b), excitons are called spin-bright if electron and hole assume a parallel spin-configuration and consequentially own a non-vanishing optical transition dipole moment. On the contrary, excitons with anti-parallel spin-configuration are called spin-dark, due to a vanishing optical transition dipole moment, and thus can not be seen in the absorption spectrum. 

The manipulation of excitonic spin properties in atomically thin semiconductors magnetic fields is a fascinating and ongoing subject of research: 
Out-of-plane magnetic fields induce Zeeman- and diamagnetic shifts \cite{wang2015magneto,forste2020exciton,article:THEORY_magnetic_Stroucken2021,delhomme2025magneticbrighteninglightlikeexcitons} and excitonic Landau levels at high field-strengths \cite{article:EXP_diamagnetic_shift_high_field_GorycaCrooker2019,stier2016exciton,van2025landau}.  In-plane magnetic fields induce spin coupling of spin-bright and spin-dark excitonic transitions via the spin angular momentum, cf.\ \cref{fig:setting}(a)--(b), which leads to the optical brightening of former spin-forbidden $s$-orbital states in photoluminescence (PL) or absorption \cite{scharf2017magnetic,vasconcelos2018dark,Van_der_Donck_2018,Feierabend_2020,Lu_2020,Robert_2020,Zhang_2017,PhysRevB.102.115420}.

In a similar way, in-plane electric fields yield substantial Stark shifts of optically bright $s$-orbital states \cite{haastrup2016stark,pedersen2016exciton,scharf2016excitonic} and cause an angular-momentum mixing \cite{zhu2023plane} in absorption spectra. On the other hand, out-of-plane electric fields cause a weak quantum-confined Stark effect \cite{engel2019electric} and spin coupling via Rashba spin-orbit interaction \cite{book:Spin_orbit_coupling_Winkler2003}, which leads to the optical brightening of mixed spin-forbidden $s$- and $p$-orbital states \cite{cao2024emergent,cao2025tunable}.

In this paper, we employ a thorough analytical approach to the effects of in-plane magnetic fields on the energetically lowest excitons in atomically thin semiconductors. We present full analytical expressions for the optical absorption, hybridized energies, linewidths and amplitudes including dissipative processes such as radiative decay or phonon-assisted decay. The analytical formulas can be easily used to analyze experimental spectra. 
We distinguish between two classes of materials: spin-dark materials -- such as MoS$_2$, WS$_2$ and WSe$_2$ -- , where the spin-dark state lies energetically below the spin-bright state; in this case, their energetic separation  exceeds usually 10 meV \cite{Robert_2020, doi:10.1021/acs.jpclett.3c02431,Molas_2017}. On the other hand, spin-bright materials -- such as MoSe$_2$ -- exhibit a spin-bright state that lies energetically below the spin-dark state; these states are energetically separated by roughly 1\,meV \cite{Lu_2020, Robert_2020}.
The paper is organized as follows: First, we develop the equations of motion for the spin-bright and spin-dark excitonic dipole (\cref{subsec: Equation of motion}) and their solutions  (\cref{subsec: Solution frequency domain}) under the influence of an in-plane magnetic field, then, we calculate the absorption spectra of two example TMDC monolayers of the two different classes, MoSe$_2$ and MoS$_2$ (\cref{sec: linear absorption spectrum}), and, at last, we analytically analyze the occurring magnetic-field-dependent spectral features (\cref{subsec: Magnetic excitonic energy}--\cref{subsec: Hybridized amplitude ratios}).

\section{\label{subsec: Equation of motion} Equations of Motion}

To construct the absorption $\alpha(\omega)$ within the coupled Maxwell-Bloch formalism \cite{KNORR199627}, we need the macroscopic polarization $P^{\, \sigma}$ induced by an external optical field $E_0^{\, \sigma}$ with polarization $\sigma = \sigma_{\pm}$, cf.\ \cref{fig:setting}(c). Since we consider an optical field, which strikes the sample perpendicularly, cf.\ \cref{fig:setting}(c), we restrict ourselves to intravalley transitions with vanishing in-plane center-of-mass momentum. In rotating-wave approximation, $P^{\, \sigma}$ reads:
\begin{multline}
        P^{\, \sigma}(t) =   \sum_{\xi} (\delta_{\sigma, \sigma_+} \delta_{\xi,K} + \delta_{\sigma, \sigma_-} \delta_{\xi,K'} ) \\   \times \sum_{s, \nu, \textbf{q}} \varphi_{\nu, \textbf{q}}^{\xi,s,s}(d_{\xi,s}^{\, c,v})^* P^{\, \xi,s,s}_{\nu}(t).
        \label{eq: macroscopic interband polarization}
\end{multline}
Here, $P^{\, \xi,s,s}_{\nu}$ is the excitonic transition \cite{https://doi.org/10.1002/pssb.201800185}:
\begin{equation}
\label{eq: excitonic transition}
    P^{\, \xi,s_1,s_2}_{\nu} = \sum_{\mathbf q}\varphi_{\nu, \textbf{q}}^{* \, \xi, s_1,s_2} \langle \hat{v}^{\dagger,\xi,s_1}_{\mathbf q} \hat{c}^{\xi,s_2}_{\mathbf q}\rangle,
\end{equation}
where $\mathbf q$ are the relative momenta of the corresponding electron-hole pair with valence (conduction) band creation (annihilation) operator $\hat{c}^{\dagger}$ ($\hat{v}$) at valley $\xi$ and spin $s$. $\nu$ is the excitonic quantum number and $\varphi_{\nu, \mathbf{q}}^{\xi,s_1,s_2}$ is the excitonic wave function solving the Wannier equation and $d_{\xi,s}^{\, c,v}$ is the absolute value of the transition dipole moment \cite{dipole}.

The exciton dynamics is obtained via Heisenberg's equations of motion:
\begin{equation}
\label{eq: Heisenberg equation of motion}
    \partial_t P^{\, \xi,s_1,s_2}_{\nu} = \frac{i}{\hbar}\langle \big[ \hat{H}, \ \hat P^{\, \xi,s_1,s_2}_{\nu} \big] \rangle,
\end{equation}
in the regime of linear optics.
The total Hamiltonian
\begin{equation}
\hat H =  \hat H_{\text{X-0}} + \hat{H}_{\text{X-light}} + \hat{H}_{B_{\parallel}},
\end{equation}
consists of the free excitonic Hamiltonian $\hat{H}_{\text{X-0}}$, the exciton-light interaction Hamiltonian in length gauge $\hat{H}_{\text{X-light}}$ and the contribution due to a spatially homogeneous in-plane magnetic field interacting with the spin angular momentum $\hat{H}_{B_{\parallel}}$:
\begin{multline}
    \hat H_{B_{\parallel}} = \sum_{\substack{\nu_1,\nu_2,\\\xi,s_1,s_2}}\left(\mathcal B_{\nu_1,\nu_2}^{e,\xi,s_1,s_2}\hat P^{\dagger}\vphantom{P}_{\nu_1}^{\xi,s_1,s_2}\hat P\vphantom{P}_{\nu_2}^{\xi,s_1,\bar s_2}\right.\\
    \left.- \mathcal B_{\nu_1,\nu_2}^{h,\xi,s_1,s_2}\hat P^{\dagger}\vphantom{P}_{\nu_1}^{\xi,s_1,s_2}\hat P\vphantom{P}_{\nu_2}^{\xi,\bar s_1, s_2}\right).
\end{multline}
We neglect diamagnetic contributions \cite{stier2016exciton,PhysRevB.98.075438} for the molybdenum-based materials for an in-plane geometry because they are sufficiently small (roughly 0.03 meV at 30\,T according to our computations) such that they are not of major importance -- even though it causes subtle asymmetries for the level repulsion in Fig. 4 -- and it is beyond the scope of this paper.
The in-plane interaction strength is governed by the excitonic magnetic matrix elements:
\begin{align}
\label{eq: matrix element}
\begin{split}
    \mathcal{B}_{\nu_1,\nu_2}^{e,\xi, s_1, s_2} &=  \frac{g_e}{2}\mu_B B_\parallel \sum_\textbf{q} \varphi_{\nu_1, \mathbf{q}}^{*\,\xi, s_1, s_2} \varphi_{\nu_2, \textbf{q}}^{\xi, s_1,\bar s_2}, \\
    \mathcal{B}_{\nu_1,\nu_2}^{h, \xi, s_1, s_2} &= \frac{g_e}{2}\mu_B B_\parallel \sum_\mathbf{q} \varphi_{\nu_1, \mathbf{q}}^{*\,\xi, s_1, s_2} \varphi_{\nu_2, \mathbf{q}}^{\xi, \bar s_1,s_2},
    \end{split}
\end{align}
where $\mu_B = \frac{e\hbar}{2m_0}$ is the Bohr magneton and $B_{\parallel}$ is the magnetic field strength. The $g$-factor can approximately be described by the free-electron $g$-factor $g_e \approx 2$ \cite{PhysRevX.4.011034, Lu_2020, PhysRevB.102.115420} and are not valley-dependent or influenced by Berry-curvature contributions due to the in-plane geometry \cite{srivastava2015valley,xiao2010berry}. 
Note that $\overline{s}$ corresponds to the opposite spin of $s$, i.e., if $s = \, \uparrow$, then $\overline{s} = \, \downarrow$ and vice versa, which is caused by the action of an in-plane magnetic field on the spin wave functions. Therefore, the magnetic Hamiltonian is non-diagonal with respect to the out-of-plane spin states of the electrons and holes of the corresponding excitons. The consequences of this situation for linear optical spectra are the main focus of this paper. The magnetic matrix elements in \cref{eq: matrix element} determine the spin-coupling selection rules: Due to the spatial homogeneity of the in-plane magnetic field $B_{\parallel}$, a coupling of angle-independent $s$-orbital states with angle-dependent $p$-orbital states does not occur. Moreover, a coupling between the $1s$-exciton and higher-lying $ns$-excitons quickly fades as $n$ increases due to a decreasing overlap of the corresponding exciton wave functions. Note that the former is a characteristic property of the interaction of a spatially homogeneous in-plane magnetic field with the spin angular momentum, as excitonic spin flips via spin-orbit interaction within spatially homogeneous \cite{cao2024emergent,mittenzwey2025ultrafast} or spatially inhomogeneous \cite{mittenzwey2025many} out-of-plane electric fields, i.e., Rashba interaction \cite{book:Spin_orbit_coupling_Winkler2003}, behave differently. Lastly, we note that due to the spin-orbit coupling, which is effectively incorporated in the excitonic Hamiltonian $\hat{H}_{\text{X-0}}$ and diagonal in the out-of-plane spin basis \cite{kormanyos2014spin,kormanyos2013monolayer,Kormanyos_2015}, the excitonic Hamiltonian does not contain any contributions from in-plane spins, which leads to an in-plane rotational spin invariance. Additionally, the spin part is not affected by the lattice geometry because the Hamiltonian factorizes the Bloch part -- which encodes the geometry -- and the spin part. As of such, different in-plane field-orientations (e.g. zig-zag or armchair) yield equivalent results. \\

By using bosonic excitonic commutator relations \cite{https://doi.org/10.1002/pssb.201800185} valid in the limit of linear optics, we obtain the following equations of motion for the excitonic transitions:

\begin{align}
\label{eq: Differential equation}
\begin{split}
 \partial_t P_{\nu_1}^{\, \xi, s_1, s_2}(t) = &\,
- \Big( \frac{i}{\hbar}\varepsilon_{x,\nu_1}^{\xi,s_1, s_2}  + \gamma_{\nu_1}^{\xi, s_1, s_2}\Big) P_{\nu_1}^{\,\xi, s_1, s_2}(t)\\
&+ i \Omega_{\nu}^{\xi, s_1, s_2}(t)
\\
& - \frac{i}{\hbar} \sum_{\nu_2} \mathcal{B}_{\nu_1, \nu_2}^{e,\xi, s_1, s_2} P_{\nu_2}^{\, \xi,  s_1,\bar s_2}(t)
\\
& + \frac{i}{\hbar} \sum_{\nu_2} \mathcal{B}_{\nu_1, \nu_2}^{h, \xi, s_1, s_2} P_{\nu_2}^{\, \xi, \bar s_1,  s_2}(t) .
\end{split}
\end{align}

The first line of \cref{eq: Differential equation} denotes the free excitonic contribution with excitonic energy $\varepsilon_{x, \nu_1}^{\xi, s_1, s_2}$  and total dephasing:
\begin{align}
    \gamma_{\nu}^{\xi, s_1, s_2} = \gamma_{\text{nrad},\nu}^{\xi, s_1, s_2} + \gamma_{\text{rad},\nu}^{\xi, s_1, s_2},
\end{align}
which consists of a non-radiative dephasing $\gamma_{\text{nrad}}^{\xi, s_1, s_2}$ with a contribution due to exciton-phonon interaction, which can be calculated and differs for respective monolayer materials due to different excitonic landscapes \cite{Selig_2016}, or due to disorder \cite{PhysRevB.68.035316} and strain \cite{Khatibi_2018}, which are introduced phenomenologically, and a radiative contribution $ \gamma_{\text{rad},\nu}^{\xi, s_1, s_2} = \frac{\varepsilon_{x,\nu}^{\xi,s_1,s_2}}{\hbar^2 2\epsilon_0c_0 n_{\text{refr}}
}|d_{\xi, s_1}^{\, c,v}|^2\delta_{s_1,s_2}$ due to reradiation \cite{KNORR199627} with vacuum permittivity $\epsilon_0$, speed of light $c_0$ and mean refractive index $n_{\text{refr}} = \frac{1}{2}(\sqrt{\epsilon_1} + \sqrt{\epsilon_2})$ of the homogeneous media surrounding the TMDC monolayer from both sides. Note that we assume $\epsilon_1=\epsilon_2$ throughout this work. The second line describes the coupling to an optical field via the excitonic Rabi frequency:
\begin{align}
\label{eq: excitonic Rabi energy}
\begin{split}
   \Omega_{\nu}^{\xi, s_1, s_2}(t) = &\, \frac{1}{\hbar}\sum_{\textbf{q}} \varphi_{\nu, \textbf{q}}^{*\,\xi,s_1, s_2} d_{\xi, s_1}^{\, c,v} \\& \times \sum_\sigma ( \delta_{\xi, K}\delta_{\sigma,\sigma_+} +  \delta_{\xi, K'}\delta_{\sigma,\sigma_-}) E_0^{\sigma}(t) ,
   \end{split}
\end{align}
which depends on the dipole matrix element $d^{\, c,v}_{\xi, s_1}$, the incident optical field $E_0^{\sigma}(t)$ with polarization $\sigma$ and the Coulomb enhancement $\sum_{\textbf{q}} \varphi_{\nu, \textbf{q}}^{*\,\xi,s_1, s_2}$. The third and fourth line of \cref{eq: Differential equation} describes the coupling of an excitonic transition with quantum number $\nu_1$ to all other transitions with quantum number $\nu_2$ via electron spin flips (third line) and hole spin flips (second line) due to the effect of the in-plane magnetic field $B_\parallel$.

\section{\label{subsec: Solution frequency domain}Hybridized Excitonic States}
Without loss of generality, in the following, we restrict ourselves to right-handed polarization $\sigma = \sigma_+$ that can optically excite electron-hole pairs at the $K$-valley only. Since we work in the regime of linear optics, our description is equally valid for left-handed or linear optical excitation and we can neglect intervalley coupling mechanisms, which are mostly relevant in nonlinear dynamics \cite{selig2019ultrafast,PhysRevLett.124.257402,deckert2025coherent,berghauser2018inverted,schmidt2016ultrafast,PhysRevB.104.245434,dogadov2025dissectingintervalleycouplingmechanisms} or states with non-vanishing center-of-mass momenta \textbf{Q} \cite{PhysRevLett.115.176801,PhysRevB.89.205303}.
Moreover, we restrict the description to the lowest-lying $1s$-excitons of the A-series, 
where we neglect A/B-coupling, since it is strongly suppressed due to the large valence band splitting of several hundreds of meV \cite{Kormanyos_2015}. The remaining two equations of motion from \cref{eq: Differential equation} can be solved analytically in frequency space by first eliminating the spin-dark ($d$) excitonic transition in the emerging analytical expression of the spin-bright ($b$) excitonic transition. The spin-bright transition $P^b(\omega) = P_{1s}^{K,\uparrow,\uparrow}(\omega)$, which couples directly to the Maxwell field via \cref{eq: macroscopic interband polarization} and contains -- due to spin mixing -- also spectral signatures of the dark transition, then reads:
\begin{align}
\label{eq: solution frequency}
        P^{b}(\omega) &= \frac{\hbar\Omega^b(\omega) \big(\hbar \omega_{x}^d - \hbar \omega - i \hbar \gamma^{d} \big)}{\big(\hbar \omega_{x}^b - \hbar \omega - i \hbar \gamma^{b}\big) \big( \hbar \omega_{x}^d - \hbar \omega - i \hbar \gamma^{d} \big) -  \mathcal{B}^2   }
\end{align}
Here, \mbox{$b = \{K, \uparrow, \uparrow, 1s \}$} denotes \textit{spin-bright} and \mbox{$d = \{K, \downarrow, \uparrow, 1s \}$} denotes \textit{spin-dark} with respect to the initial excitonic configuration at $B_{\parallel} = 0$, cf.\ \cref{fig:setting}(a)--(b) and we define $\mathcal{B} = |\mathcal{B}_{ \nu}^{e,K, \uparrow, \uparrow}|$, cf.\ \cref{eq: matrix element}.
Subsequently, we can cast the equation in a more intuitive form by performing a partial fraction decomposition:

\begin{align}
\label{eq: solution summarized}
P^b(\omega) = \hbar\Omega^b(\omega)\sum_{\mathcal S}   \frac{P^{\mathcal S}_{ B_\parallel}}{\hbar \omega_{x,B_\parallel}^{\mathcal S} - \hbar \omega - i\hbar \gamma_{ B_\parallel}^{\mathcal S}}.
\end{align}
As a consequence of the magnetic-field induced spin coupling, the spin-bright transition in \cref{eq: solution summarized} is now a superposition of 
two excitonic resonances, the hybridized states labeled by the spin-diagonal quantum number $\mathcal S \in \{1, -1 \} $ centered around their respective hybridized energies:
\begin{multline}
\label{eq: magnetic frequency}
\hbar \omega_{x, B_\parallel}^{\mathcal S} =  \frac{1}{2} (\hbar \omega_{x}^b + \hbar \omega_{x}^d) \\ +\mathcal S \frac{1}{2\sqrt{2}} \sqrt{ \mathcal{C}(\mathcal{B};\Delta;\kappa) - \kappa^2  + \Delta^2 + 4\mathcal{B}^2},
\end{multline}
with hybridized linewidths:
\begin{multline}
\label{eq: magnetic dephasing}
\hbar \gamma_{ B_\parallel}^{\mathcal S} = \frac{1}{2} (\hbar \gamma^d + \hbar \gamma^b)\\
+\mathcal S \frac{\text{sgn}(\Phi)}{2\sqrt{2}}   \sqrt{ \mathcal{C}(\mathcal{B};\Delta;\kappa) + \kappa^2  - \Delta^2 - 4\mathcal{B}^2},
\end{multline}
We define:
\begin{multline}
     \mathcal{C}(\mathcal{B};\Delta;\kappa) =  \\  \sqrt{8 \mathcal{B}^2\Big( - \kappa^2 + \Delta^2 + 2 \mathcal{B}^2 \Big) + \Big( \kappa^2 + \Delta^2 \Big)^2},
\end{multline}
with dark-bright splitting $\Delta$:
\begin{align} 
\label{eq: dark bright splitting}
    \Delta = |\hbar  \omega_{x}^b - \hbar \omega_{x}^d|,
\end{align}
and linewidth difference $\kappa$:
\begin{align}
\label{eq: linewidth difference}
    \kappa = | \hbar \gamma^b - \hbar \gamma^d|.
\end{align}
The quantity:
\begin{equation}
\label{eq: Phi}
    \Phi =  (\omega^{b}_{x} - \omega^{d}_{x}) (\gamma^b - \gamma^d),
\end{equation}
determines the sign in \cref{eq: magnetic dephasing} and ensures correct convergence behavior in the limit $B_\parallel \rightarrow 0$.

The amplitudes of both occurring resonances in \cref{eq: solution summarized} are scaled with the complex mixing coefficients $P^{\mathcal S}_{B_\parallel}$:
\begin{align}
\label{eq: mixing coefficient}
P^{\mathcal S}_{ B_\parallel} = \frac{\omega_{x}^d - i \gamma^d  - \omega_{x,B_\parallel}^{\mathcal S} + i \gamma_{ B_\parallel}^{\mathcal S} }{ \omega_{x,B_\parallel}^{\overline{\mathcal S}} - i \gamma_{ B_\parallel}^{\overline{\mathcal S}}  -  \omega_{x,B_\parallel}^{\mathcal S} + i \gamma_{ B_\parallel}^{\mathcal S}},
\end{align}
which obey the following identity
\begin{equation}
\label{eq: conservation dipole omment}
P^{\mathcal S}_{ B_\parallel} + P^{\overline{\mathcal S}}_{ B_\parallel} = 1.
\end{equation}
To obtain a form that reveals the dependencies of the mixing coefficient transparently -- while losing the compactness of the expression -- , we can insert the definitions \cref{eq: magnetic frequency,eq: magnetic dephasing} and transform to polar coordinates; from this, we learn that the mixing coefficient does not parametrically depend on individual linewidths nor energies but only on the dark-bright splitting $\Delta$ and linewidth difference $\kappa$. 

We note that due to the explicit inclusion of different linewidths of the involved spin-bright and spin-dark excitons, which directly impact the spin hybridization, our excitonic description resembles a non-Hermitian treatment to the optical response of atomically thin semiconductors \cite{wang2024non}. These non-Hermiticy-resembling effects are most pronounced, if energy splitting, linewidth and linewidth differences between the involved resonances are all of similar magnitude. Such a regime can occur in, e.g., disorder-/defect-free h-BN-encapsulated MoSe$_2$ monolayers at cryogenic temperatures with a dark-bright energy splitting $\Delta$ of 1.5\,meV and a linewidth $\hbar\gamma$ of the spin-bright exciton of approximately $1$--$1.5\,$meV.

\begin{figure}
    \centering
    \includegraphics[width=1\linewidth]{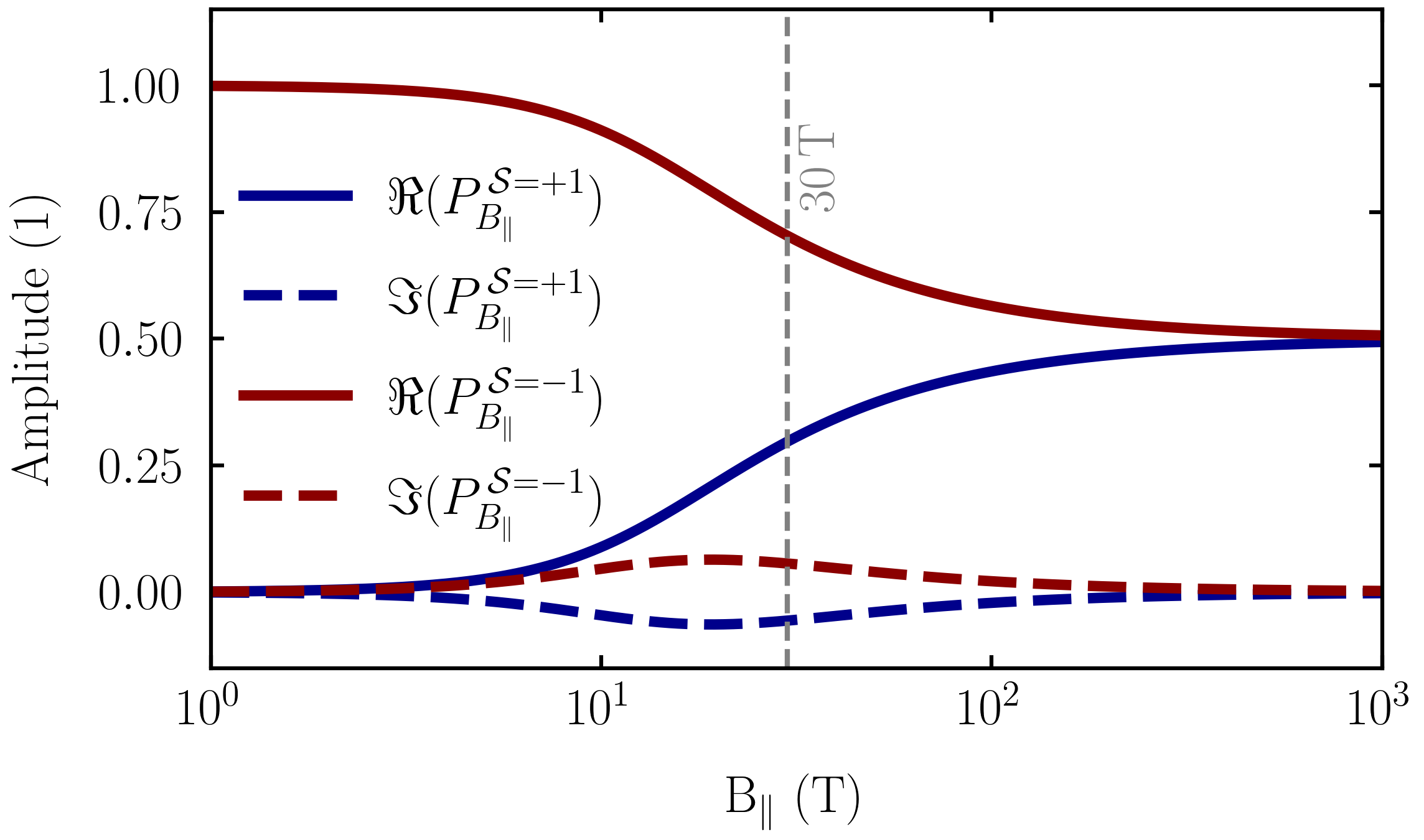}
    \caption{Real- and imaginary part of mixing coefficients $P_{B_\parallel}^{\mathcal S}$ for $\nu = 1s$ under influence of the in-plane magnetic field for MoSe$_2$  ($\kappa = 0.5$ meV).}
    \label{fig: mixing coefficient}
\end{figure}

The mixing coefficients from \cref{eq: mixing coefficient} are plotted as an example for small non-radiative linewidth in \cref{fig: mixing coefficient}. Here, the real part redistributes the oscillator strength with increasing magnetic field between both initial spin-bright and spin-dark excitonic transitions and therefore displays the degree of spin hybridization: At zero magnetic field, no redistribution takes place and the mixing coefficient corresponding to the spin-dark excitonic transition is zero. At an increasing magnetic field, the mixing coefficient of the initial spin-bright transition decreases and the mixing coefficient of the initial spin-dark transition increases, until both mixing coefficients converge to
\begin{equation}
\lim_{B_\parallel \mapsto \infty} P^{\mathcal S}_{ B_\parallel} = \lim_{B_\parallel \mapsto \infty} P^{\overline{\mathcal S}}_{ B_\parallel} = 0.5
\end{equation}
for very large magnetic field strengths. However, this regime clearly exceeds currently experimentally available field strengths on the order of 100 T \cite{article:EXP_diamagnetic_shift_high_field_GorycaCrooker2019}. Even without an in-plane magnetic field, intrinsic spin-hybridization at the $K$ valleys \cite{deilmann2020ab,junior2022first} is present. However, in molybdenum-based materials, this spin-hybridization is much less than 10\,\%. Hence, within our model, we can safely assume an ideal spin model, since the spin-hybridization induced by in-plane magnetic fields at 30\,T -- cf. Fig. 2 -- assumes values around 30\,\% that greatly exceeds the intrinsic spin-hybridization.
In \cref{eq: mixing coefficient} and \cref{fig: mixing coefficient}, additional imaginary components appear because we do not neglect dissipation throughout our whole analysis. 

Lastly, we note that, due to varying energy ordering of spin-bright and spin-dark excitons in the respective sample of interest, the assignment of the $\mathcal S$-states to specific spin-bright or spin-dark states can change.

\section{Linear Absorption Spectrum \label{sec: linear absorption spectrum}}

The linear absorption spectrum $\alpha(\omega)$ 
can be expressed as \cite{bookKnorr}:
\begin{equation}
\label{eq: absorption}
    \alpha(\omega) = 1 - T(\omega) - R(\omega),
\end{equation}
where $T(\omega)$ and $R(\omega)$ are the transmission and reflection coefficients, respectively:
\begin{align}
    T(\omega) &= \bigg| \frac{\textbf{E}_{T}(\omega)}{\textbf{E}_0(\omega)} \bigg|^2 = \Bigg| \frac{\Tilde{E}_0^{\sigma_+}(\omega) + \frac{i\varepsilon^b_x}{2 \hbar \epsilon_0 c_0 n_{\text{refr}}} \Tilde{P}^{\sigma_+}(\omega)}{\Tilde{E}^{\sigma_+}_0(\omega)} \Bigg|^2, \\
R(\omega) &= \bigg| \frac{\textbf{E}_{R}(\omega)}{\textbf{E}_0(\omega)} \bigg|^2 = \Bigg| \frac{ \frac{i\varepsilon^b_x}{2 \hbar \epsilon_0 c_0 n_{\text{refr}}} \Tilde{P}^{\sigma_+}(\omega)}{\Tilde{E}^{\sigma_+}_0(\omega)} \Bigg|^2 ,
\end{align}
that are computed with the macroscopic polarization $P^\sigma(t)$ from \cref{eq: macroscopic interband polarization}. By performing a partial fraction decomposition again, making a close-to-resonance approximation (discussed in App. \ref{app: absorption spectrum}) and with the help of \mbox{\cref{eq: solution summarized}}, the absorption spectrum can be expressed explicitly as:

\begin{align}
\label{eq: analytical absorption spectrum}
    \alpha(\omega) = \sum_{\mathcal S}  \frac{\mathcal{A}^{\mathcal{S}}_{B_\parallel}}{(\hbar \omega_{x,B_\parallel}^{\mathcal S} - \hbar \omega)^2 + (\hbar \gamma_{ B_\parallel}^{\mathcal S} )^2},
\end{align}
with the spin-hybridized amplitudes:
\begin{align}
\begin{split}
\label{eq: amplitude}
    \mathcal{A}^{\mathcal{S}}_{B_\parallel}  &= -2  (\hbar \gamma_{\text{rad}}^b)^2|P^\mathcal{S}_{ B_\parallel}|^2 \\&\quad +  2\hbar \gamma_{ B_\parallel}^\mathcal{S} \big( \hbar \gamma_{\text{rad}}^b \text{Re} [ P^\mathcal{S}_{ B_\parallel}]    + 2 \mathcal{S} \text{Im} [\Lambda_{B_\parallel}] \big),
\end{split}
\end{align}
which is a superposition of the two peak signals with amplitudes $\mathcal{A}^{\mathcal{S}}_{B_\parallel}$ that effectively incorporate interference effects: The first term on the right hand side of \cref{eq: amplitude} comprises the amplitude -- present without applying an in-plane magnetic field -- that is redistributed among the two spin-hybridized resonances; the second term originates from interference between the transmitted field and the incident optical field; the third term that scales with $\Lambda_{B_\parallel}$ (defined in \cref{app: absorption spectrum}) arises from interference between the two spin-hybridized resonances. This formula provides a motivation for experimentalists to fit their spectra with a pair or Lorentzians. Its derivation is discussed in App. \ref{app: absorption spectrum}.

\begin{figure}[h]
    \centering
    \includegraphics[width=1\linewidth]{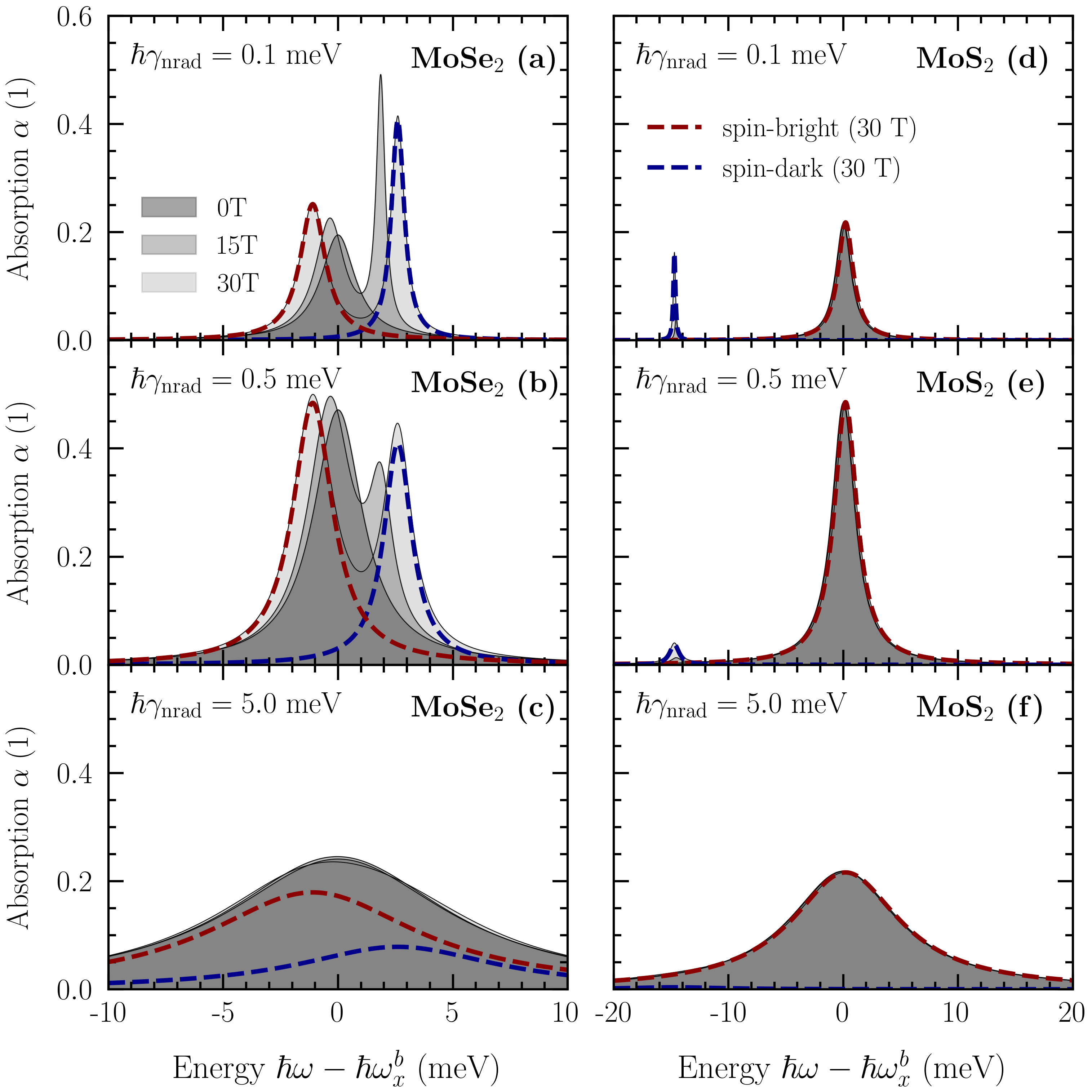}
    \caption{Linear absorption spectrum under influence of an in-plane magnetic field for $\nu = 1s$ with different field-strengths for the material MoSe$_2$ in (a)--(c) and MoS$_2$ in (d)--(f) encapsulated in h-BN with $\sigma_+$-polarised light for increasing non-radiative linewidths $\hbar \gamma_{\text{nrad}}$ as tuning parameter}. 
    \label{fig: spectrum MoSe2}
\end{figure}

We depict in \cref{fig: spectrum MoSe2} the linear absorption spectra \cref{eq: analytical absorption spectrum} of monolayers MoSe$_2$ (left) and MoS$_2$ (right) encapsulated in h-BN for three in-plane magnetic field-strengths around the spin-bright 1s-orbital state. From top to bottom, we vary the non-radiative linewidth $\hbar \gamma_{\text{nrad}}^{b/d} = \hbar \gamma_{\text{nrad}} $, which describes the regime of a  linewidth that is dominated by reradiation up to a regime, where phonons gain a dominating influence (100\,K) 
\cite{Selig_2016}. Additionally, we display the two individual resonance components ($\mathcal{S}=1$ and $\mathcal{S}=-1$) from \cref{eq: analytical absorption spectrum} at 30\,T. We chose an h-BN-embedding because it enables high-purity samples with comparably small linewdiths \cite{shree2018observation,cadiz2017excitonic,ajayi2017approaching}.

To obtain \cref{fig: spectrum MoSe2}, we adjusted the excitonic energies according to recent PL experiments with dark-bright splitting of 1.45\,meV in the optically bright MoSe$_2$ monolayer \cite{Lu_2020,Robert_2020} and dark-bright splitting of 14.5\,meV in the optically dark MoS$_2$ monolayer \cite{Robert_2020}. We calculated the Coulomb enhancement $\sum_\textbf{q} \varphi_{1s, \textbf{q}}^{* \, K, \uparrow, \uparrow}$ in \cref{eq: excitonic Rabi energy} mediated by the excitonic wave functions as solutions of the corresponding Wannier equations and the transitions dipole moments with $\mathbf k\cdot\mathbf p$-parameters according to Refs.~\cite{dipole,Kormanyos_2015}. All parameters are provided in \cref{tab:parameters} in the appendix.

In the absorption of a MoSe$_2$ monolayer in \cref{fig: spectrum MoSe2} at increasing in-plane magnetic field strengths, the initially spin-dark resonance brightens, i.e., gains an oscillator strength. Simultaneously, both excitonic resonances shift in the opposite direction, i.e., an anti-crossing occurs. This has also been observed in PL \cite{Robert_2020, Feierabend_2021, Lu_2020, D0NR04243A, Molas, Zhang_2017}. In the MoS$_2$ monolayer, the same behavior occurs, but strongly suppressed due to the much larger dark-bright splitting of 14.5\,meV compared to 1.45\,meV in the MoSe$_2$ monolayer. The brightening of initially spin-dark excitons is most pronounced with a negligible non-radiative linewidth compared to the radiative linewidth -- cf. \cref{fig: spectrum MoSe2}(a) --, where a strong signal at the spin-dark resonance already appears at 15\,T. At a slightly increased non-radiative linewidth of $\hbar \gamma_{\text{nrad}} = 0.5 $ meV, a clear spin-dark resonance appears at 30\,T and a shoulder can be observed at 15\,T in a MoSe$_2$ monolayer, while, in a MoS$_2$ monolayer, the initial spin-dark resonance is barely detectable even at 30\,T. At comparably large non-radiative linewidths of $\hbar \gamma_{\text{nrad}} = 5\,$meV, both resonances can no longer be separated and appear as one single, broadened peak. Therefore, we require small dark-bright splitting for strong spin-hybridization and sufficiently non-overlapping resonances such that the spin-hybridization remains optically detectable.

We observe an enhanced amplitude -- even without in-plane magnetic field -- when raising the non-radiative linwidth from $\hbar \gamma_{\text{nrad}} = 0.1$\,meV to $\hbar \gamma_{\text{nrad}} = 0.5$. This is due to stronger interference between the initial optical field and the spin-hybridized resonances (second term from \cref{eq: amplitude}), which leads to enhanced absorption. The amplitude drops for $\hbar \gamma_{\text{nrad}} = 5.0$\,meV because a further increase of amplitude is canceled by broadening. We note that normalized values (w.r.t. spin-bright amplitude without magnetic field and corresponding non-radiative linewidth) of the spin-hybridized amplitude $\mathcal{A}^\mathcal{S}_{B_\parallel}$ show a negligible dependence on the non-radiative linewidth. This implies that, while the non-radiative linewidth governs the absolute magnitudes of the spin-hybridized amplitudes $\mathcal{A}_{B_\parallel}^{\mathcal{S}}$, it does not affect the global spectral form. Also, difficult to detect by eye, but present, is a redistribution of the linewidths in an increasing magnetic field, which is modelled by \cref{eq: magnetic dephasing}. Also, difficult to detect by eye, but present, is a redistribution of the linewidths in an increasing magnetic field, which is modelled by \cref{eq: magnetic dephasing}.

These three magnetic-field-dependent features -- the peak position, the linewidth and the behavior of the amplitude -- are discussed in detail in the following.

\subsection{Hybridized Excitonic Energies}
\label{subsec: Magnetic excitonic energy}
\begin{figure}[h]
    \begin{subfigure}[b]{1\linewidth}
        \includegraphics[width=1\linewidth]{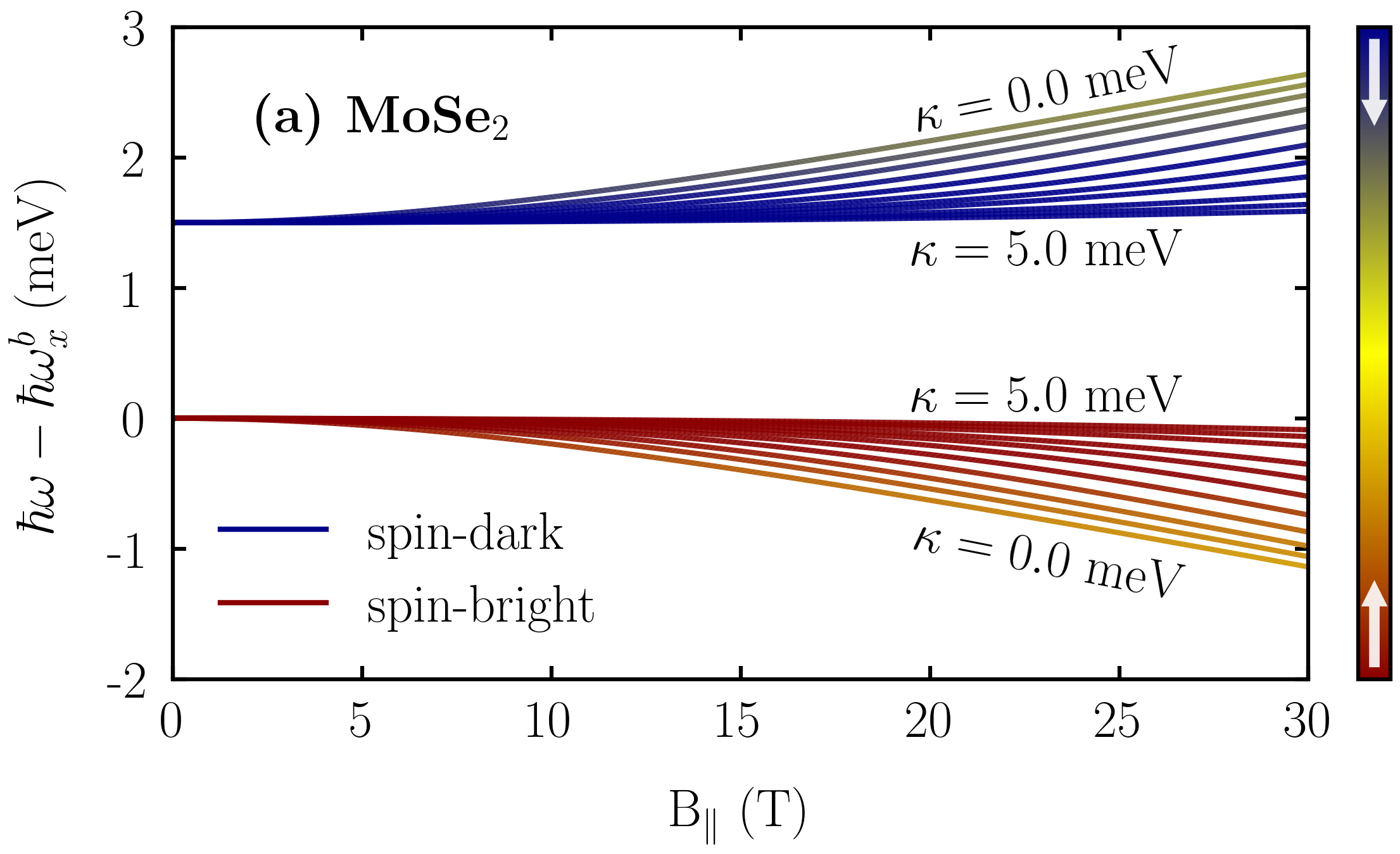}
        \caption{MoSe$_2$}
        \label{fig: energies MoSe2}
    \end{subfigure}
    
    \vspace{0cm} 

    \begin{subfigure}[b]{1\linewidth}
    \includegraphics[width=1\linewidth]{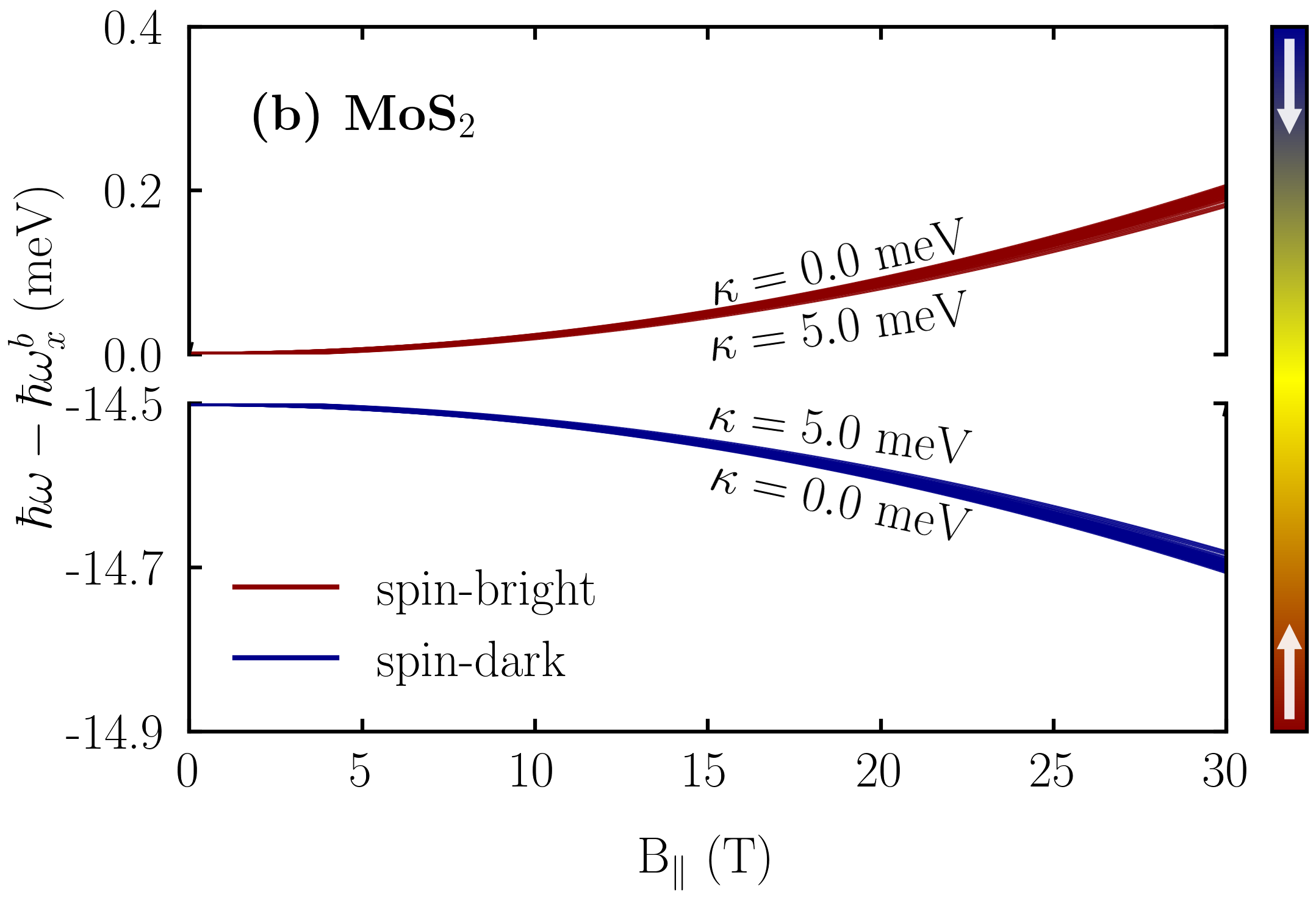} 
        \caption{MoS$_2$}
    \label{fig: energies MoS2}
    \end{subfigure}
    \caption{Hybridized excitonic energies $\hbar \omega_{x,B_\parallel}^\mathcal{S}$ from \cref{eq: magnetic frequency} for MoSe$_2$ in \cref{fig: energies}(a) and MoS$_2$ in \cref{fig: energies}(b) under influence of an in-plane magnetic field for increasing linewidth difference $\kappa$ from \cref{eq: linewidth difference}. The colourgradient of the curves show the values of the real parts of mixing coefficients from \cref{eq: mixing coefficient}, i.e. the hybridization coefficients. Red (blue) represents a spin-up (spin-down) electron-state and yellow denotes the degree of spin-mixing.}
    \label{fig: energies}
\end{figure}

The hybridized excitonic energies $\hbar \omega_{x,B_\parallel}^\mathcal{S}$  are governed by \cref{eq: magnetic frequency} and plotted for the materials MoSe$_2$ in \cref{fig: energies MoSe2} and MoS$_2$ in \cref{fig: energies MoS2} as a function of magnetic field $B_\parallel$ and linewidth difference $\kappa$ from \cref{eq: linewidth difference}.

As expected, the hybridized excitonic energies repel each other under influence of the in-plane magnetic field. In all cases, the initial growth as a function of $B_\parallel$ is quadratic. This can be seen analytically by performing a Taylor approximation at $B_\parallel = 0$, yielding
\begin{equation}
\label{eq: magnetic frequency taylor}
     \hbar \omega_{x, B_\parallel}^{\mathcal{S}} \approx  \frac{1}{2} (\hbar \omega_{x}^b + \hbar \omega_{x}^d) + \mathcal{S}  \Big(\frac{1}{2} \Delta \, + \mathcal{B}^2   \frac{\Delta}{\kappa^2 + \Delta^2} \Big).
     \end{equation}
Here, for small  $B_\parallel$-fields, the excitonic hybridized frequency grows quadratically in the field $B_{\parallel}$ and linear in the dark-bright splitting $\Delta$, renormalized by both the dark-bright splitting and the linewidth difference. The monolayer MoS$_2$ in \cref{fig: energies}(b) does not leave this quadratic regime, because the excitonic magnetic matrix element still remains relatively small compared to the dark-bright splitting even up to 30 T.

On the contrary, for very high magnetic-field strengths and small dark-bright splittings, the growth becomes linear, which can be observed in the case for MoSe$_2$ in olive \cref{fig: energies}(a). This regime can be analytically obtained from \cref{eq: magnetic frequency}, if one assumes that the in-plane magnetic field dominates:
\begin{equation}
        \hbar \omega_{x,B_\parallel}^\mathcal{S} \approx \frac{1}{2 } (\hbar \omega_{x}^b + \hbar \omega_{x}^d) + \mathcal{S}  \mathcal{B}.
\end{equation}

In the following, we discuss the impact of the linewidth difference on the level repulsion. The dependence on the linewidth difference is particularly pronounced for MoSe$_2$ in \cref{fig: energies}(a), because it exhibits a small dark-bright splitting. As indicated by the color gradient, the spin-hybridization at a fixed magnetic field strength varies, if the linewidth difference is varied: If $\kappa$ increases, the degree of hybridization and hence the splitting decreases. Therefore, the dissipation of the system effectively suppresses spin-hybridization and can also be viewed as an effective attractive interaction.

To illustrate this situation, we distinguish two analytical limit cases. (i) In the case for dominating linewidth differences, we can eliminate the dark-bright-splitting-dependency by assuming that $\kappa \gg \Delta$ in the model \cref{eq: magnetic frequency taylor}, yielding:
\begin{equation}
    \hbar \omega_{x, B_\parallel}^\mathcal{S} \approx \frac{1}{2} (\hbar \omega_{x}^b + \hbar \omega_{x}^d) + \mathcal{S}  \Big( \frac{1}{2} \Delta \, + \mathcal{B}^2 \frac{\Delta}{\kappa^2} \Big).
\end{equation}
In this case, the initial quadratic growth becomes exclusively inversely proportional to the square of $\kappa$. This holds true for MoSe$_2$ in \cref{fig: energies}(a) for dominating $\kappa$. (ii) On the other hand, when the dark-bright splitting $\Delta$ becomes the dominating parameter, which holds true for the majority of the TMDCs (MoS$_2$, WS$_2$, WSe$_2$), we obtain an expression that does not depend on $\kappa$ anymore:
\begin{align}
\label{eq: shifts delta dominates}
\hbar \omega_{x, B_\parallel}^\mathcal{S}  \approx  \frac{1}{2} (\hbar \omega_{x}^b + \hbar \omega_{x}^d) + \mathcal{S}  \Big(\frac{1}{2} \Delta \, + \frac{\mathcal{B}^2}{\Delta} \Big),
\end{align}
i.e., the influence of the linewidth difference vanishes. This regime is valid for MoS$_2$ in \cref{fig: energies}(b), where the linewidth difference $\kappa$ only influences the shifts of energy marginally. At last, we emphasize the limit of $\kappa \approx 0$:
\begin{align}
\hbar \omega_{x, B_\parallel}^\mathcal{S} = \frac{1}{2 } (\hbar \omega_{x}^b +  \hbar \omega_{x}^d) +  \frac{\mathcal{S}}{2  } \sqrt{ \Delta^2 + 4 \mathcal{B}^2}
\label{eq:usual_level_repulsion_inplaneB}
\end{align}
This case, for both materials in \cref{fig: energies}(a)--(b), leads to the strongest shifts, as the attractive part of the coupling induced by the dissipation is absent. \cref{eq:usual_level_repulsion_inplaneB} is the usual expression of level repulsion \cite{sakurai2017modern} and has been already discussed in other works \cite{Lu_2020, Feierabend_2020}. In \cref{tab:table}, we summarize the measured level repulsion at 15\,T and 30\,T and compare these values with our model according to \cref{eq: magnetic frequency}; we observe a good agreement between experiment and theory. We suspect that small differences at 30\,T are due to varying linewidth differences $\kappa$ for both experiments.

\begin{table}[h]
    \centering
    \begin{tabular}{c|c|c|c|c|c|}
        Ref. & Material & $B_\parallel = 15$\,T (Exp./Th.) & $B_\parallel = 30$\,T (Exp./Th.)  \\
    \hline
         \cite{Robert_2020} & MoSe$_2$ & 0.4 / 0.4 meV & 1.2 / 1.06 meV  \\
         \cite{Lu_2020} & MoSe$_2$ & 0.4 / 0.4 meV & 1.1 / 1.06 meV  \\
    \end{tabular}
    \caption{Level repulsion of experiments (Exp.) and our theory at $\kappa = 0$ (Th.) at 15\,T and 30\,T.}
    \label{tab:table}
\end{table}
 We note that the magnetic-field-induced hybridization also manifests in time domain via quantum beats \cite{Mapara2022}.
 
\subsection{Hybridized Excitonic Linewidths}
\label{subsec: Hybridized excitonic dephasing}
\begin{figure}[h]
    \centering
        \includegraphics[width=\linewidth]{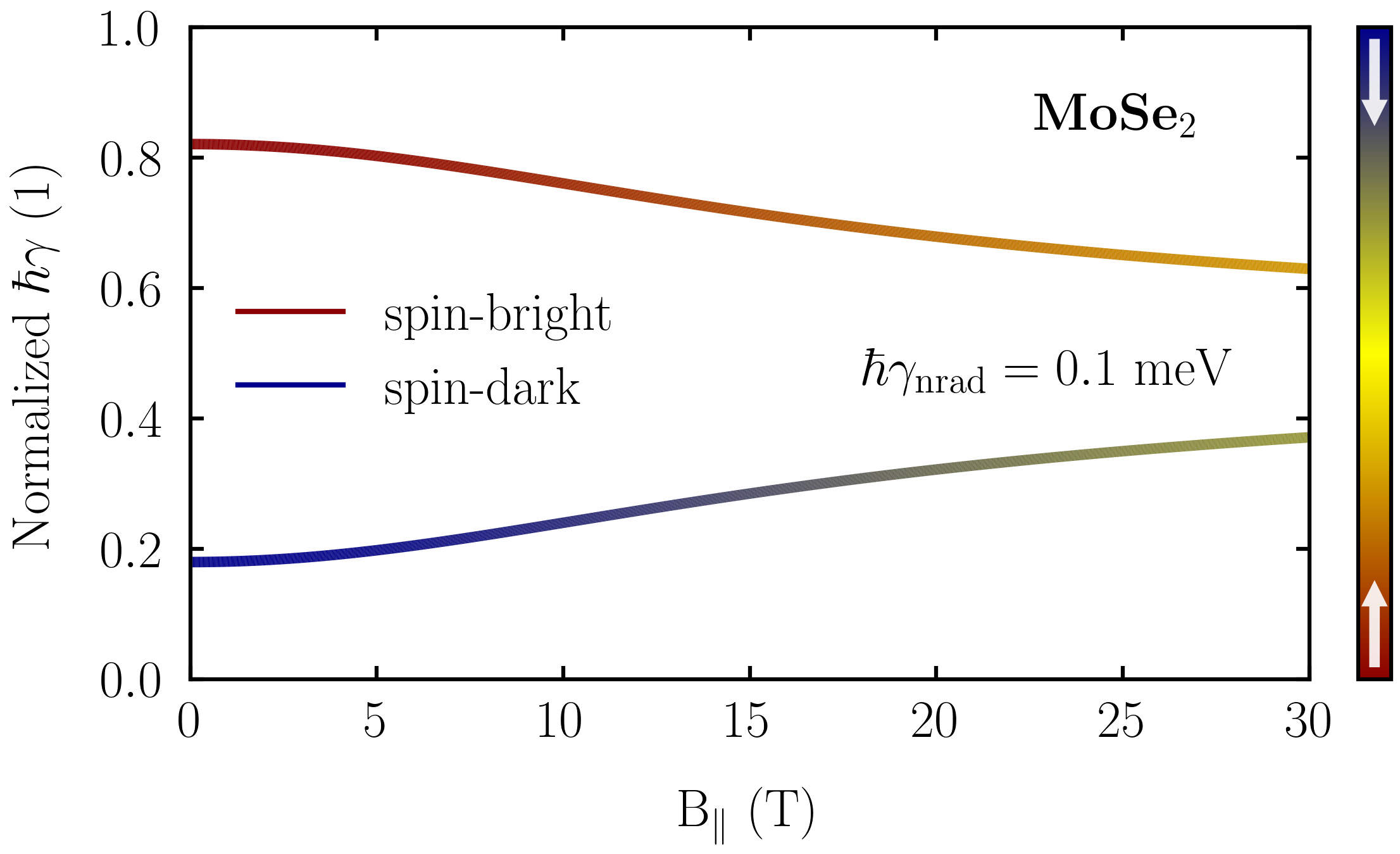}
        \caption{Hybridized linewidth $\hbar \gamma_{ B_\parallel}^\mathcal{S}$, normalized w.r.t. summed linewidth $\hbar \gamma^b + \hbar \gamma^d$,  in dependence of an in-plane magnetic field for MoSe$_2$ with $\kappa$ = 0.71 meV and $\hbar \gamma_{\text{nrad}} = 0.1$ meV. The colourgradient of the curves show the values of the  real parts of mixing coefficients from \cref{eq: mixing coefficient}, i.e. the hybridization coefficients. Red (blue) represents a spin-up (spin-down) electron-state  and yellow denotes the degree of spin-mixing.}
        \label{fig:linewidth_a}
\end{figure}

The hybridized linewidths $\hbar \gamma_{ B_\parallel}^\mathcal{S} $ are governed by \cref{eq: magnetic dephasing} and plotted for MoSe$_2$ in \cref{fig:linewidth_a}. We note that the linewidths, discussed in this manuscript, always correspond to a coherence decay via, e.g., radiative or pure dephasing and are \textit{not} equal to a population decay due to radiative or nonradiative recombination \cite{selig2018dark}.

\begin{figure}
    \centering
\includegraphics[width=1.0\linewidth]{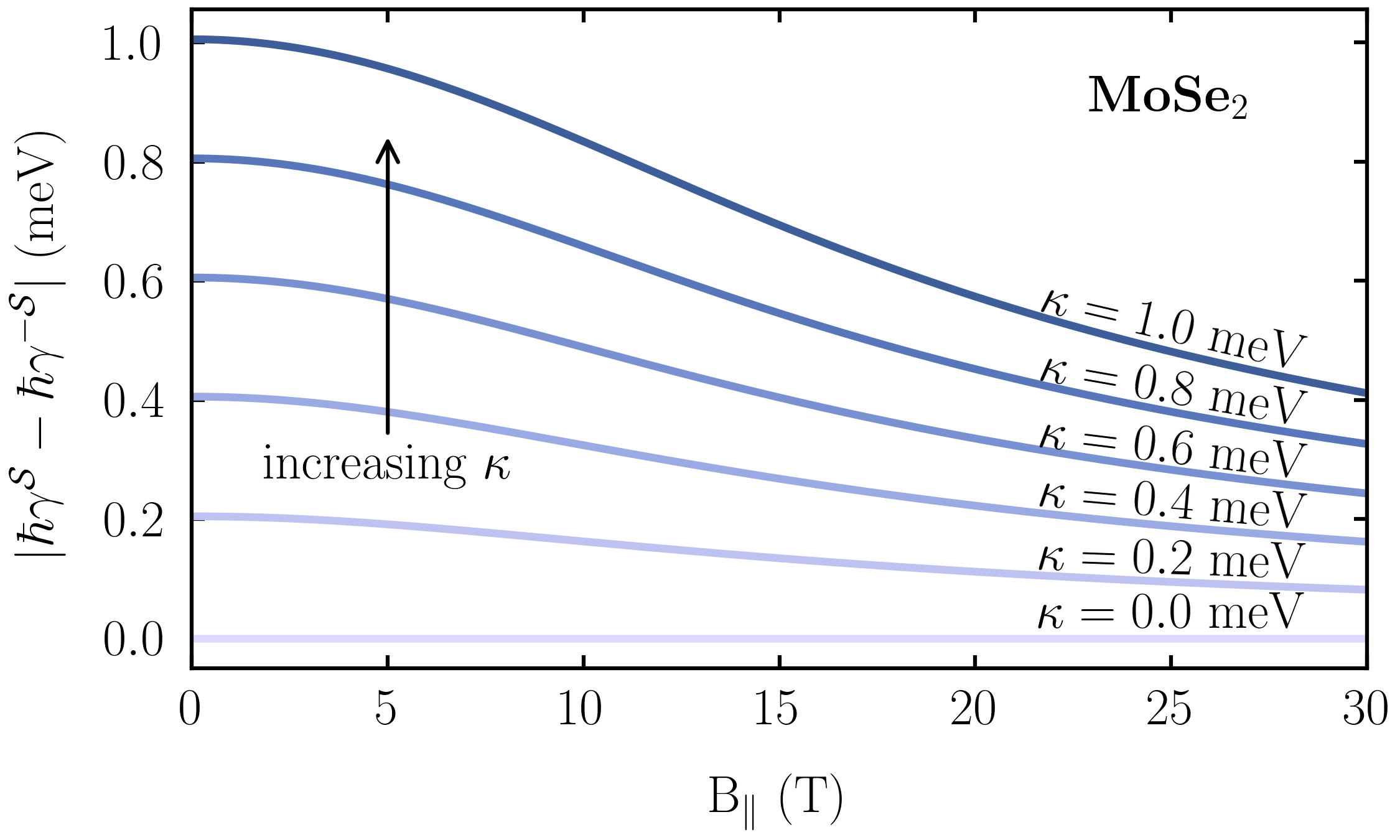}
        \caption{Absolute difference of hybridized linewidths calculated with \cref{eq: magnetic linewidth difference} for MoSe$_2$ under influence of of an in-plane magnetic field for increasing linewidth differences $\kappa$}
        \label{fig:linewidth_b}
\end{figure}

At increasing magnetic fields, it can be seen from \cref{fig:linewidth_a} that the linewidths of the two hybridized states grow towards each other, which results in the fact -- in contrast to the magnetic-field-dependent behavior of the hybridized excitonic energies, cf.\ \cref{fig: energies} -- that the hybridized linewidths approach their mean value at very high magnetic fields: 
\begin{equation}
    \lim_{B_\parallel \mapsto \infty}  \hbar \gamma_{ B_\parallel}^\mathcal{S} = \frac{1}{2}( \hbar \gamma^b + \hbar \gamma^d).
\end{equation}
In general, the resonance with the larger linewidth distributes a fraction to the resonance with the smaller linewidth. To demonstrate this in more detail, we depict the absolute difference of the hybridized linewidths  in \cref{fig:linewidth_b}:
\begin{multline}
\label{eq: magnetic linewidth difference}
 |\hbar \gamma_{ B_\parallel}^\mathcal{S} - \hbar \gamma_{ B_\parallel}^{\overline{\mathcal{S}}}| =\frac{1}{\sqrt{2}}   \sqrt{ \mathcal{C}(\mathcal{B};\Delta;\kappa) + \kappa^2  - \Delta^2 - 4\mathcal{B}^2}.
\end{multline}
For vanishing field-strength, it holds:
\begin{align}
\label{eq: magnetic linewidth difference square root}
|\hbar \gamma_{ B_\parallel}^\mathcal{S} - \hbar \gamma_{ B_\parallel}^{\overline{\mathcal{S}}}| \Big|_{B_\parallel = 0} &= \kappa,
\end{align}
cf.\ \cref{fig:linewidth_b} at $B_{\parallel} = 0\,$T.  Therefore, if no linewidth difference is present, i.e.\ $\kappa = 0$, there is no linewidth to redistribute and the impact of the magnetic field on the individual linewidths vanishes. Analytically, this follows from \cref{eq: magnetic dephasing} directly via:
\begin{equation}
    \hbar \gamma_{ B_\parallel}^\mathcal{S} \Big|_{\kappa = 0} = \hbar \gamma^b = \hbar \gamma^d,
\end{equation}
which is depicted in \cref{fig:linewidth_b} for the $\kappa = 0$ - case. Furthermore, for increased linewidth differences, \cref{fig:linewidth_b} highlights that the redistribution of linewidth becomes stronger in terms of absolute values, yet needs stronger fields to converge to the mean value as the hybridization coefficients grows weaker as discussed in \cref{subsec: Solution frequency domain}. Similarly to \cref{subsec: Magnetic excitonic energy}, further limiting cases of \cref{eq: magnetic dephasing} can be constructed, whose derivation we omit here.

As in \cref{fig: energies}, the hybridized linewidths in a MoSe$_2$ monolayer, cf.\ \cref{fig:linewidth_a}, are much more sensitive to the magnetic field compared to MoS$_2$, WS$_2$ and WS$_2$ monolayers due to its smaller dark-bright splitting.

\subsection{Absorption ratios of hybridized exciton peaks}
\label{subsec: Hybridized amplitude ratios}
\begin{figure}[h]
    \centering

    \begin{subfigure}[b]{1\linewidth}
        \centering
        \includegraphics[width=\linewidth]{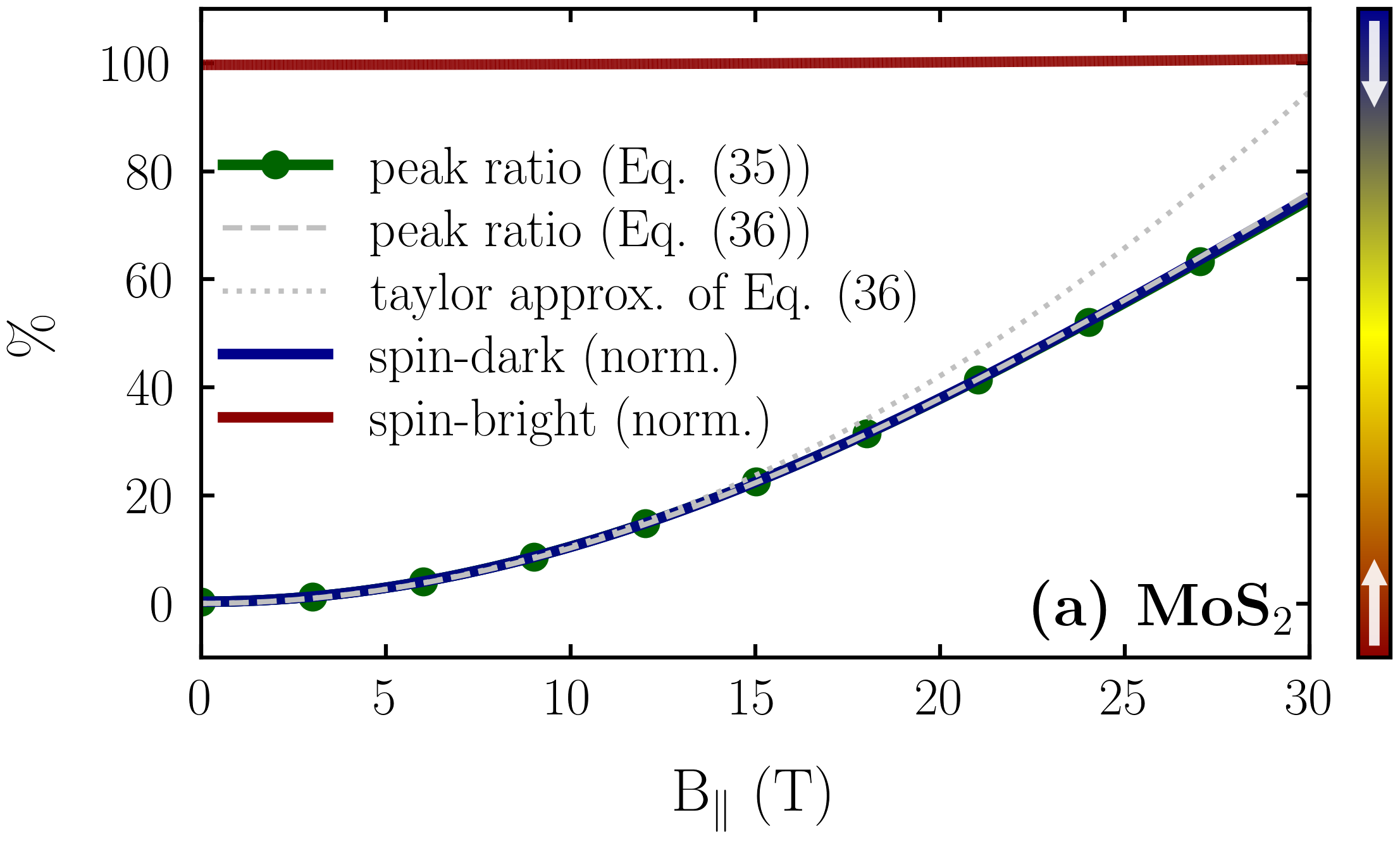}
        \caption{MoS$_2$}
        \label{fig:amplitude_ratio_MoS2}
    \end{subfigure}

        \begin{subfigure}[b]{1\linewidth}
        \centering
        \includegraphics[width=\linewidth]{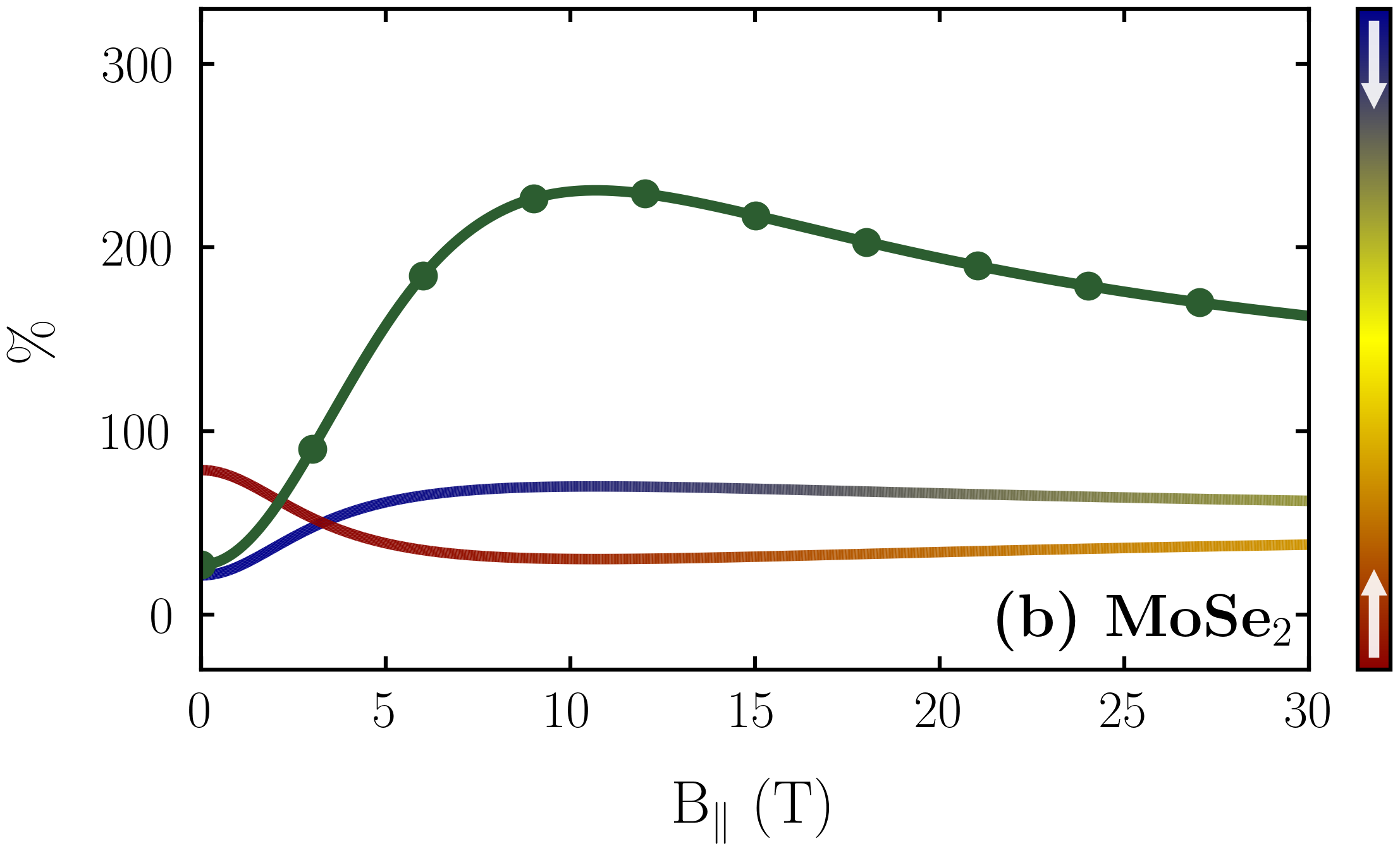}
        \caption{MoSe$_2$}
        \label{fig:amplitude_ratio_MoSe2}
    \end{subfigure}
    \caption{Spin-bright/spin-dark amplitudes, that are normalized w.r.t. $\mathcal{N} = \alpha(\hbar \omega_{x, B_\parallel = 0}^\mathcal{S}) + \alpha(\hbar \omega_{x, B_\parallel = 0}^{\overline{\mathcal{S}}})$ and amplitude peak ratios  for MoS$_2$ ( \cref{fig:amplitude_ratios}(a)) and MoSe$_2$ (\cref{fig:amplitude_ratios}(b)) with non-radiative linewidth $\hbar \gamma_{\text{nrad}} = 0.1$ meV for increasing in-plane magnetic field-strengths. The colourgradient of the curves show the values of the  real parts of mixing coefficients from \cref{eq: mixing coefficient}, i.e. the hybridization coefficients. Red (blue) represents a spin-up (spin-down) electron-state  and yellow denotes the degree of spin-mixing. }
    \label{fig:amplitude_ratios}
\end{figure}

By evaluating the absorption spectrum in \cref{eq: analytical absorption spectrum} at the hybridized excitonic energies \cref{eq: magnetic frequency}, we obtain the amplitude peak ratio of spin-dark and spin-bright resonances as:
\begin{align}
\label{eq: relative amplitudes}
\begin{split}
    &\text{Peak~ratio} = \frac{\alpha(\hbar \omega_{x, B_\parallel}^\mathcal{S})}{\alpha(\hbar \omega_{x, B_\parallel}^{\overline{\mathcal{S}}})} .
    \end{split}
\end{align}
In  \cref{fig:amplitude_ratios}(a) and \cref{fig:amplitude_ratios}(b), we show the peak amplitude ratios from \cref{eq: relative amplitudes} for the materials MoS$_2$ and MoSe$_2$, respectively. We evaluate \cref{eq: relative amplitudes} in such a way that the numerator (denominator) always corresponds to the spin-dark (spin-bright) resonance, and we choose a regime where the radiative and non-radiative linewidths are roughly on the same scale, which corresponds to disorder-/defect-free h-BN-encapsulated samples at cryogenic temperatures.
In \cref{fig:amplitude_ratios}(a), the peak ratio of MoS$_2$ (olive solid line) displays a monotonous increase within a magnetic field strength interval of up to 30$\,$T. This behavior reflects the increased optical brightening of the initial spin-dark exciton, cf.\ also \cref{fig: energies}(b) and \cref{fig: mixing coefficient}.
We can derive an analytical limit case for \cref{eq: relative amplitudes} by utilizing the fact that the dark-bright splitting dominates over the linewidths. 
We evaluate numerator and denominator separately by using \cref{eq: shifts delta dominates} and a similar limiting case for \cref{eq: magnetic dephasing} and keeping only the terms that grow quadratically in the magnetic field strength and find:
\begin{align}
\begin{split}
    \label{eq: peak ratio MoS2}
&\text{Peak~ratio} \, \Bigg|_{\text{MoS}_2} \approx \\ &\frac{(\hbar \gamma^b)^2}{(\hbar \gamma^d )^2} \frac{\mathcal{B}^2}{(\hbar \gamma^b - \hbar \gamma_{\text{rad}}^b) \Delta^2 + [\hbar \gamma^d +2\frac{\hbar \gamma^b}{\hbar \gamma^d}(\hbar \gamma^b - \hbar \gamma_{\text{rad}}^b)] \mathcal{B}^2}.
\end{split}
\end{align}
This approximate expression, cf.~silver dashed line in \cref{fig:amplitude_ratios}(a), describes the initial grow with respect to the magnetic-field strength, which reproduces the full expression, cf.~olive solid line, very well. For magnetic fields above 30 T, however, \cref{eq: peak ratio MoS2} fails to reproduce the correct saturation behavior: It asymptotically approaches a ratio exceeding  100\,$\%$, whereas the exact expression exhibits a maximum before converging to 100\,$\%$ as illustrated for MoSe$_2$ in \cref{fig:amplitude_ratios}(b). We further plot the quadratic approximation of \cref{eq: peak ratio MoS2}, cf.~silver dotted line in \cref{fig:amplitude_ratios}(a), with respect to the excitonic magnetic matrix element $\mathcal{B}^2$ around the origin, 
which reproduces the full expression (olive solid line) up to roughly 15 T. 
We note that the limit case in \cref{eq: peak ratio MoS2} can be used to model any material that exhibits a dominating dark-bright splitting over the linewidths, for example WS$_2$ and WSe$_2$ at cryogenic temperatures.

In contrast, within the field strengths considered here, the peak amplitude ratio for a MoSe$_2$ monolayer, cf.~olive solid line in \cref{fig:amplitude_ratios}(b), displays a strong non-monotonous behavior at increasing magnetic-field strengths. Initially, it increases, until a turning point at around 10\,T is reached, from where it then slowly decreases.
This non-monotonous behavior is a direct consequence of the fact that in a MoSe$_2$ monolayer, dark-bright splitting, linewidths and linewidths differences are all of similar magnitude in the range of 1--2\,meV. 
The strong initial growth of the peak ratio well beyond 100~$\%$ can be explained by the fact that, due to the absence of a radiative contribution, the linewidth of the spin-dark resonance is smaller compared to the spin-bright resonance. This in turn overemphasizes the oscillator strength it receives from hybridization (smaller linewidths lead to larger peak amplitudes at a constant area under the curve). For a closer analysis, we rewrite the peak ratio from \cref{eq: relative amplitudes} as:
\begin{align}
\label{eq: amplitudes MoSe2}
\begin{split}
    &\text{Peak~ratio} \Big|_{\text{MoSe}_2} = \frac{(\hbar \gamma_{ B_\parallel}^{\mathcal{S}=-1})^2}{(\hbar \gamma_{ B_\parallel}^{\mathcal{S}=1})^2}  g^{\mathcal{S } = 1},  
    \end{split}
\end{align}
which is a product of a monotonously decreasing ratio of the hybridized linewidths $\frac{(\hbar \gamma_{ B_\parallel}^{\mathcal{S}=-1})^2}{(\hbar \gamma_{ B_\parallel}^{\mathcal{S}=1})^2}$ and a monotonously increasing function $g^\mathcal{S}$ with respect to the magnetic field $B_{\parallel}$, defined as
\begin{align}
\begin{split}
&g^\mathcal{S} =  \frac{\big[(\hbar \omega_{x,B_\parallel}^{\mathcal S} - \hbar \omega_{x, B_\parallel}^{\overline{\mathcal{S}}})^2 + (\hbar \gamma_{ B_\parallel}^{\mathcal S} )^2\big]}{\big[(\hbar \omega_{x,B_\parallel}^{\overline{\mathcal S}} - \hbar \omega_{x, B_\parallel}^\mathcal{S})^2 + (\hbar \gamma_{ B_\parallel}^{\overline{\mathcal{S}}} )^2 \big]} \\
    & \times\frac{ \mathcal{A}^{\mathcal{S}}_{B_\parallel}\big[(\hbar \omega_{x,B_\parallel}^{\overline{\mathcal{S}}} - \hbar \omega_{x, B_\parallel}^\mathcal{S})^2 + (\hbar \gamma_{ B_\parallel}^{\overline{\mathcal{S}}} )^2\big] +\mathcal{A}^{\overline{\mathcal{S}}}_{B_\parallel} (\hbar \gamma_{ B_\parallel}^{\mathcal S} )^2}{\mathcal{A}^{\overline{\mathcal{S}}}_{B_\parallel} \big[(\hbar \omega_{x,B_\parallel}^{\mathcal S} - \hbar \omega_{x, B_\parallel}^{\overline{\mathcal{S}}}]^2 + (\hbar \gamma_{ B_\parallel}^{\mathcal S} )^2\big] + \mathcal{A}^{\mathcal{S}}_{B_\parallel}(\hbar \gamma_{ B_\parallel}^{\overline{\mathcal{S}}} )^2 }.
    \end{split}
    \label{eq:g_function}
\end{align}
Hence, the maximum of the amplitude ratio in  \cref{fig:amplitude_ratios}(b) can be rationalized as that point, where the linewidth ratio $\frac{(\hbar \gamma_{ B_\parallel}^{\mathcal{S}=-1})^2}{(\hbar \gamma_{ B_\parallel}^{\mathcal{S}=1})^2}$ reflecting the magnetic-field-induced linewidth redistribution starts to grow faster than the oscillator-strength redistribution encoded in $g^\mathcal{S}$.  At magnetic fields exceeding the turning point, a decrease of the amplitude ratio occurs, until it is fully determined by the degree of the spin hybridization, as $\frac{(\hbar \gamma_{ B_\parallel}^{\mathcal{S}=-1})^2}{(\hbar \gamma_{ B_\parallel}^{\mathcal{S}=1})^2}\overset{B_{\parallel}\rightarrow \infty}{\longrightarrow} 1$ and $g^\mathcal{S}\overset{B_{\parallel}\rightarrow \infty}{\longrightarrow}$ 1. 
Consequently, if the linewidths are initially equal, the linewidth ratio becomes unity and the the peak ratio in \cref{eq: amplitudes MoSe2} will be a purely monotonic increase with no extremum, as in the case for MoS$_2$ in \cref{eq: peak ratio MoS2} (within the field strengths considered). Note that within our definition, cf.~\cref{eq: relative amplitudes}, the spin-dark amplitude and, thus, the ratio are not exactly zero for a vanishing in-plane magnetic field, cf.~ \cref{fig:amplitude_ratios}(b), because the spin-dark resonance lies in close vicinity to the spin-bright resonance. 

We point out that the qualitative behavior of the peak ratios resembles the behavior of the peak ratios extracted from PL experiments \cite{Robert_2020,Lu_2020}. However, while similar in their outer appearance, the underlying physics is different: In contrast to absorption measurements, which display the coherent response of the material, PL displays the emission of radiation. Hence, the PL amplitudes always scale with the incoherent excitonic occupations at the respective exciton resonance, which recombine within the light cone \cite{selig2018dark,brem2020phonon,Feierabend_2021, Molas_2017}, and the peak ratio can be rationalized as the ratio of the Boltzmann-distributed spin-bright and spin-dark occupations \cite{Lu_2020}. Therefore, future work 
can possibly shed more light on this issue, i.e., on the question, if the non-monotonous behavior can also be observed in actual absorption measurements.

\section*{Conclusion}
We derived a fully analytical model for the absorption spectrum of the excitonic spin-dark and spin-bright ground states under spin hybridization of an applied in-plane magnetic field. We found, that an intricate interplay of hybridized excitonic energies, linewidths and amplitudes governs the linear optical response. In particular, we showed that a MoSe$_2$ monolayer with sufficiently small linewidths gives rise to a non-monotonous amplitude behavior of spin-bright and spin-dark excitonic states in optical absorption spectra, similar to recent photoluminescence experiments. The derived equations \cref{eq: analytical absorption spectrum,eq: magnetic frequency,eq: magnetic dephasing}, can be used to analyze magneto-optical experiments in the field of TMDC excitons.

\section*{Data availability statement}
All data that support the findings of this study are included within the article.
\acknowledgments
The authors acknowledge financial support by the Deutsche Forschungsgemeinschaft (DFG) through Project No. 420760124. H.M.\ acknowledges funding by project 21209528 (``proof of trust'').

\appendix

\section{Absorption spectrum}
\label{app: absorption spectrum}
In this appendix, we show how we arrive at the absorption formula in \cref{eq: analytical absorption spectrum} by introducing a Lorentzian of the spin-hybridized state $\mathcal S$ as
\begin{align}
\label{eq: def resonance}
L_\mathcal{S}(\omega) &= 
P^\mathcal{S}_{ B_\parallel}  \frac{i \hbar \gamma_{\text{rad}}^b}{ \hbar \omega_{x,B_\parallel}^\mathcal{S} - \hbar \omega  - i\hbar \gamma_{ B_\parallel}^\mathcal{S}}.
\end{align}
With this definition, we can express the absorption spectrum from \cref{eq: absorption} as follows:
\begin{align}
\begin{split}
    \alpha(\omega)
&= 1 - |L_{\mathcal{S}=1}(\omega) + L_{\mathcal{S}=-1}(\omega)|^2  \\ & \qquad - |1 + L_{\mathcal{S}=1}(\omega) + L_{\mathcal{S}=-1}(\omega)|^2 
\\
& \qquad -  4 \Re[L_{\mathcal{S}=1}(\omega) L_{\mathcal{S}=-1}^*(\omega)]. 
\end{split}
\end{align}
We plug in the definition from \cref{eq: def resonance} and reorganize the expression using partial fraction decomposition, which leads to the following absorption spectrum 
\begin{align}
\label{eq: analytical absorption spectrum appendix without approx}
    \alpha(\omega) = \sum_{\mathcal S}  \frac{\mathcal{F}^{\mathcal{S}}_{B_\parallel} + \mathcal{G}^{\mathcal{S}}_{B_\parallel}  (\hbar \omega - \hbar \omega_{x}^\mathcal{S})}{(\hbar \omega_{x,B_\parallel}^{\mathcal S} - \hbar \omega)^2 + (\hbar \gamma_{ B_\parallel}^{\mathcal S} )^2},
\end{align}
The amplitudes are given by:
\begin{align}
\label{eq: amplitudes spectrum}
\mathcal{F}^{\mathcal{S}}_{B_\parallel} & = -2  (\hbar \gamma_{\text{rad}}^b)^2|P^\mathcal{S}_{ B_\parallel}|^2 \\& \qquad +  2\hbar \gamma_{ B_\parallel}^\mathcal{S} \big( \hbar \gamma_{\text{rad}}^b \text{Re} [ P^\mathcal{S}_{ B_\parallel}]    + 2 \mathcal{S} \text{Im} [\Lambda_{B_\parallel}] \big),  \notag \\
\mathcal{G}^{\mathcal{S}}_{B_\parallel} &= 2\Re\Big[i \hbar \gamma_{\text{rad}}^b P^\mathcal{S}_{ B_\parallel}  + 2\mathcal{S} \Lambda_{B_\parallel} \Big],
\end{align}
with:
\begin{equation}
    \Lambda_{B_\parallel} =   \frac{(\hbar \gamma_{\text{rad}}^b)^2 P^{\mathcal{S}=1}_{ B_\parallel} P^{* \, \mathcal{S} = -1}_{ B_\parallel}  }{\hbar \omega_{x,B_\parallel}^{{\mathcal{S} = -1}} - \hbar \omega_{x,B_\parallel}^{\mathcal{S}=1} + i (\hbar \gamma^b + \hbar \gamma^d)}.
\end{equation}
To obtain a simplified formula for the linear absorption spectrum, we can make the following close-to-resonance approximation 
\begin{equation}
\mathcal{G}^{\mathcal{S}}_{B_\parallel} \cdot (\hbar \omega - \hbar \omega_{x}^{\mathcal{S}}) \approx 0
\end{equation}
yielding the linear absorption spectrum
\begin{align}
\label{eq: analytical absorption spectrum appendix with approx}
    \alpha(\omega) = \sum_{\mathcal S}  \frac{\mathcal{A}^{\mathcal{S}}_{B_\parallel} }{(\hbar \omega_{x,B_\parallel}^{\mathcal S} - \hbar \omega)^2 + (\hbar \gamma_{ B_\parallel}^{\mathcal S} )^2},
\end{align}
which is a superposition of a pair of Lorentzians.

\section{Parameters}
\label{app: parameters}
In \cref{tab:parameters}, we display the parameters used in all analytical (optical spectra) and numerical (Wannier equation) calculations. In particular, we use the momentum-dependent model dielectric function from Ref.~\cite{Trolle}, which is fitted to \textit{ab-initio} calculations from the Computational Materials Repository \cite{andersen2015dielectric}, and the screened Coulomb potential derived in Ref.~\cite{deckert2025coherent} in the Wannier equation for the calculation of the excitonic binding energies and wave functions. To model the screening effect of the h-BN encapsulation, we fit the respective momentum-dependent dielectric function to \textit{ab-initio} results \cite{latini2015excitons} with static dielectric constant of bulk h-BN $\epsilon_{\text{h-BN},0} = \sqrt{\epsilon_{\text{h-BN},0,\parallel}\epsilon_{\text{h-BN},0,\bot}} = 4.8$ \cite{latini2015excitons}. For the light-matter interaction, we use the optical in-plane dielectric constant of h-BN $\epsilon_{\text{h-BN},\infty,\parallel} = 4.87$ \cite{cai2007infrared}. The choice of embedding (h-BN within this manuscript) influences -- through its corresponding dielectric constant -- the excitonic binding energies, the excitonic wave functions (and therefore the overlaps and Coulomb enhancement) and the radiative linewidth.

\begin{table*}
    \centering
    \begin{tabular}{lll}
    Quantity & MoSe$_2$ & MoS$_2$\\
    \hline
        $1s$ spin-bright excitonic energy $\hbar\omega_{x,1s}^b$ & 1.639\,eV (4\,K) \cite{Robert_2020} & 1.931\,eV (4\,K) \cite{Robert_2020} \\
        Dark-bright splitting $\hbar\omega_{x,1s}^b-\hbar\omega_{x,1s}^d$ & $-1.45$\,meV \cite{Lu_2020, Robert_2020} & 14.5\,meV \cite{Robert_2020, doi:10.1021/acs.jpclett.3c02431}\\
        Effective spin-up electron mass $m_e^{K,\uparrow}$ \cite{Kormanyos_2015} & $0.5m_0$ & $0.44m_0$ \\
        Effective spin-down electron mass $m_e^{K,\uparrow}$ \cite{Kormanyos_2015} & $0.58m_0$ & $0.47m_0$ \\
        Effective spin-up hole mass $m_h^{K,\uparrow}$ \cite{Kormanyos_2015} & $0.6m_0$ & $0.54m_0$\\
        $\mathbf k \cdot \mathbf p$-parameter $\gamma_{\mathbf k\cdot \mathbf p}$ \cite{Kormanyos_2015} & $\frac{1}{2}(0.253+0.220)$\,eV\,nm & $\frac{1}{2}(0.276 + 0.222)$\,eV\,nm\\
        Transition dipole moment $d^{cv} = \frac{e\sqrt{2}\gamma_{\mathbf k\mathbf p}}{\tilde E_{\text{gap}}}$
        & 0.1689\,$e$\,nm & 0.1536\,$e$\,nm \\
        Monolayer width $d$ \cite{kylanpaa2015binding} & 0.6527$\,$nm  & $0.6180$\,nm\\
Static bulk dielectric constant 
$\epsilon_{s,0} = \sqrt{\epsilon_{s,\parallel,0}\epsilon_{s,\bot,0}}$ \cite{laturia2018dielectric} & 12.0474  & 10.4743\\
        Plasmon peak energy $\hbar\omega_{\text{pl}}$ \cite{kumar2012tunable} & 22.0\,eV  & 22.5\,eV\\
    Thomas-Fermi parameter $\alpha_{\text{TF}}$ (fit to \textit{ab-initio} results \cite{andersen2015dielectric}) & 1.9 & 1.5\\
    Interlayer gap $h$ \cite{florian2018dielectric,druppel2017diversity} & 0.3\,nm & 0.3\,nm\\
    $1s$ spin-bright excitonic binding energy (calculated) & $-341.3$\,meV & $-361.2$\,meV\\
    $1s$ spin-dark excitonic binding energy (calculated) & $-358.4$\,meV & $-369.2$\,meV\\
    Radiative linewidth $\hbar\gamma^b_{\text{rad}}$ (calculated) & 0.81\,meV & 0.71\,meV\\
    Coulomb enhancement $\sum_{\mathbf q}\varphi_{1s,\mathbf q}^{*\,K,\uparrow,\uparrow}$ (calculated) & 0.9157 & 0.8677\\
    Overlaps $\sum_{\mathbf q}\varphi_{1s,\mathbf q}^{*,K,\uparrow,\uparrow}\varphi_{1s,\mathbf q}^{K,\downarrow,\uparrow}$ (calculated) & 0.9989 & 0.9998\\
    \end{tabular}
    \caption{Parameters used in the numerical (Wannier equation) and analytical (optical spectra) calculations.}
    \label{tab:parameters}
\end{table*}

\FloatBarrier
\bibliography{bib}    

\end{document}